\documentclass{osa-article}

\journal{oe}
\usepackage{lineno}
\begin{document}

\title{Compact Plug and Play Optical Frequency Reference Device Based on Doppler-Free Spectroscopy of Rubidium Vapor}

\author{Aaron Strangfeld\authormark{1,2,*}, Benjamin Wiegand\authormark{1}, Julien Kluge\authormark{1,2}, Matthias Schoch\authormark{1}, and Markus Krutzik\authormark{1,2}}

\address{\authormark{1}Department of Physics, Humboldt-Universität zu Berlin, Newtonstraße 15, 12489 Berlin, Germany\\
\authormark{2}Ferdinand-Braun-Institut gGmbH, Leibniz-Institut für Höchstfrequenztechnik, Gustav-Kirchhoff-Str. 4, 12489 Berlin, Germany}

\email{\authormark{*}strangfa@physik.hu-berlin.de} %% email address is required



\begin{abstract*}
Compactness, robustness and autonomy of optical frequency references are prerequisites for reliable operation in mobile systems, on ground as well as in space. We present a standalone plug and play optical frequency reference device based on frequency modulation spectroscopy of the D2-transition in rubidium at 780\,nm. After a single button press the hand-sized laser module, based on the micro-integrated laser-optical bench described in [J. Opt. Soc. Am. B {\bfseries 38}, 1885-1891 (2021)], works fully autonomous and generates 6\,mW of frequency stabilized light with a relative frequency instability of 1.4$\times$10$^{-12}$ at 1\,s and below 10$^{-11}$ at 10$^5$\,s averaging time. We describe the design of the device, investigate the thermal characteristics affecting the output frequency and demonstrate short-term frequency stability improvement by a Bayesian optimizer varying the modulation parameters.
\end{abstract*}

%%%%%%%%%%%%%%%%%%%%%%%%%%  body  %%%%%%%%%%%%%%%%%%%%%%%%%%

\section{Introduction}
Atomic optical frequency references (AOFR) are critical subsystems of cold-atom based quantum sensors and optical clocks. These technologies can contribute to fundamental science \cite{NewPhysics,STEQuest}, ground- and space-based earth observation \cite{GroundEarthObs,ColdAtomsLab2Real} and navigation \cite{OptiClockNav,DSAC}. Other use cases for AOFRs are for example the calibration of optical instruments, e.g. wavelength meters \cite{Wavemeters1,Wavemeter2}, or absolute referencing of laser systems for trace-gas detection \cite{CO2Detection} or optical communication \cite{OptComm}.\\
Ideally, AOFRs serve as robust and reliable sources of frequency-stabilized laser light without further interaction with the hosting system. Functional and physical isolation is therefore advantageous allowing for improved predictability of the performance after integration. They are thus particularly well suited for realizations as standalone modules. However, the technological challenges of AOFR development are diverse, comprising the mechanical and thermal design, the integration of a reliable laser light source, the optical engineering of a spectroscopy unit and the development of suitable low-noise driving electronics. This motivates segmentation. For instance, sounding rocket missions demonstrated AOFRs in space realized by separate laser, optics and electronics subsystems \cite{Kalexus,Fokus,Jokarus} instead of one integrated module each. This design philosophy is reinforced when multiple lasers and electronics control units are required, which is the case for complex cold-atom applications \cite{CAL,BECCAL}. To allow for a transition from mission dedicated system development to flexible commercial realizations of cold-atom quantum sensors, optical clocks, wavelength meters or other laser systems, compact stand-alone AOFR modules are necessary.\\ The most advanced AOFR modules belong to the domain of warm atomic vapors. Optical clocks utilizing coherent population trapping \cite{CPT}, such as the chip scale atomic clocks \cite{ChipScaleKitching,ChipScaleProto}, are available as commercial-off-the-shelf (COTS) components \cite{MicrosemiCSAC}. Developed as integrated clocks, they generate a 10\,MHz output signal and thus can't be used directly for optical referencing of cold-atom quantum sensors or optical instruments.\\ Turn-key solutions with optical output do exist \cite{ThorlabsReference,WavelengthReference}. However, to our knowledge they are only available at telecommunication wavelengths and achieve relative frequency instabilities at the 10$^{-9}$ level at 1\,s averaging time. COTS frequency discriminator modules based on alkali-metals \cite{TopticaCosy,Vescent} can be used to achieve lower instabilities and research on a high-performing one based on modulation transfer spectroscopy (MTS) has been presented recently \cite{MTSReferenceModule}. These require a suitable external laser input and electronics control unit. Commercial laser packages with integrated reference cells \cite{Flame} leave only the development of additional electronics and software open to the user.\\
We present a standalone, hand-sized plug and play AOFR system based on the laser module we presented in \cite{Strangfeld:21} with a distributed feedback (DFB) laser diode utilizing the D2-transition of warm rubidium vapor at 780\,nm. The system achieves relative instabilities of 1.4$\times$10$^{-12}$ at 1\,s and below 10$^{-11}$ up to 10$^5$\,s averaging time. Powering it leads to autonomous thermal stabilization of the optical bench. A button press activates the laser and reliably initiates the frequency stabilization to a predefined atomic transition.\\
The work is structured as follows. Section \ref{sec:overview} gives an overview over the system components and functions with a summary of the achieved size, weight and power (SWaP). In Section \ref{sec:thermal}, the thermal characteristics such as warm-up time of the device and temperature sensitivity of the optical frequency are investigated. The utility of an automated Bayesian optimization process for the instability minimization is exemplarily demonstrated in Section \ref{sec:opti}. Section \ref{sec:perform} discusses the achieved performance in terms of the Allan deviation. Finally, in Section \ref{sec:discussion}, the results are summarized and an outlook for future developments and potential use-cases of the device is given.

\section{System Overview}\label{sec:overview}
\begin{figure}[t]
\centering
\includegraphics[width=25pc]{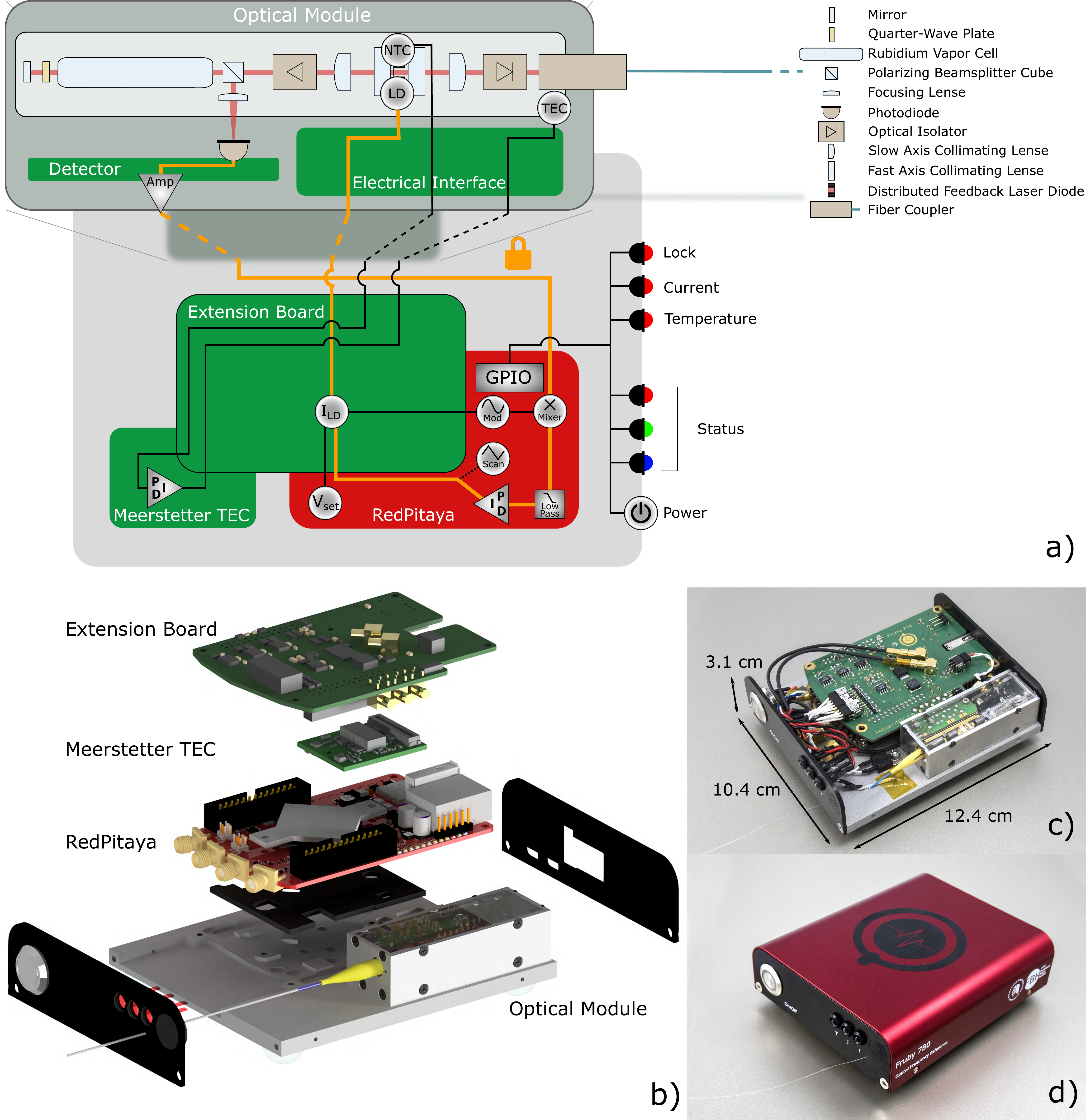}
\caption{System Overview. (a) Simplified functional scheme excluding the monitoring functions for temperature controller status, device temperature and laser diode current. (b) Explosion including the optical module, the RedPitaya StemLab, a compact temperature controller and the driver extension board for signal distribution and current driving. (c) Assembled system without cover with dimensions annotated. (d) Complete system with cover. Image Credits (c,d): P. Immerz (FBH).}
\label{fig:system_overview}
\end{figure}

The frequency reference system is built around the optical module presented in \cite{Strangfeld:21}. There, a DFB laser diode is integrated with a rubidium vapor cell and a detector into a single, temperature stabilized module for frequency modulation spectroscopy of the D2-line at 780\,nm. To realize a stand-alone system, a dedicated software and the following additional subsystems embedded in a housing are needed: a control and lock-in amplification unit, a temperature controller and a current driver. In the presented module, this is addressed by utilizing the \textit{RedPitaya StemLab 125-14} (RP) and a custom extension board, which hosts a commercial temperature controller (\textit{Meerstetter TEC-1092}) attached to the bottom side as well as an integrated current driver developed in our group based on a modified Libbrecht-Hall design \cite{LibbrechtHall}. A system overview is given in Figure \ref{fig:system_overview}.\\
To accommodate a compact design, the standard heat-sink of the RP is replaced by a flat piece of aluminum and unused connectors are removed. A thermal pad between the RP and the structured base plate ensures sufficient cooling. The system can be mounted on rubber stands and is still capable to operate with passive air cooling at room temperature. To monitor the device temperature, a thermistor is installed on the base plate. The connection to the optical module is realized by MCX-plugs soldered to the extension board. \\A software developed on the basis of \textit{Linien} \cite{linien,linien2} runs on the RP and fulfills the following tasks: monitoring of the temperature controller status, control of the current set-point, monitoring of the driving current and the device temperature, generation of the current sweep and modulation signal, demodulation of the spectroscopy signal, frequency locking and, finally, control of the front-panel LEDs. The RGB-LED integrated in the button indicates the general device status (idle, running, error). The remaining three red LEDs indicate the status of the temperature (stable/ramping), the current (on/off) and the frequency lock (locked/unlocked), respectively. A rubber cylinder is used as the feedthrough and strain relief of the polarization maintaining optical fiber. The rear panel gives access to the standard interface of the RP, comprising connectors for the power supply (micro-USB), USB type A and Ethernet.\\Table\,\ref{tab:swap} summarizes the system's SWaP as it is depicted in Figure\,\ref{fig:system_overview}. The optical output power from the polarization maintaining fiber is 6\,mW.\\ While the device works completely autonomous after the button on the front-panel is pressed, it can be connected to a network to monitor and control system parameters as well as for the selection of the transition to be locked on.
\begin{table}[htbp]
\centering
\caption{\bf Size, weight and power consumption of the frequency reference as depicted in Figure \ref{fig:system_overview} d).}
\begin{tabular}{ll}
\hline
Size & $12.4\times 10.4\times 3.1$\,cm$^3$\\
Weight & $490$\,g\\
Power & $7.0$\,W\\
\hline
\end{tabular}
 \label{tab:swap}
\end{table}
\section{Thermal Characteristics}\label{sec:thermal}
\begin{figure}
\centering
\includegraphics[width=23pc]{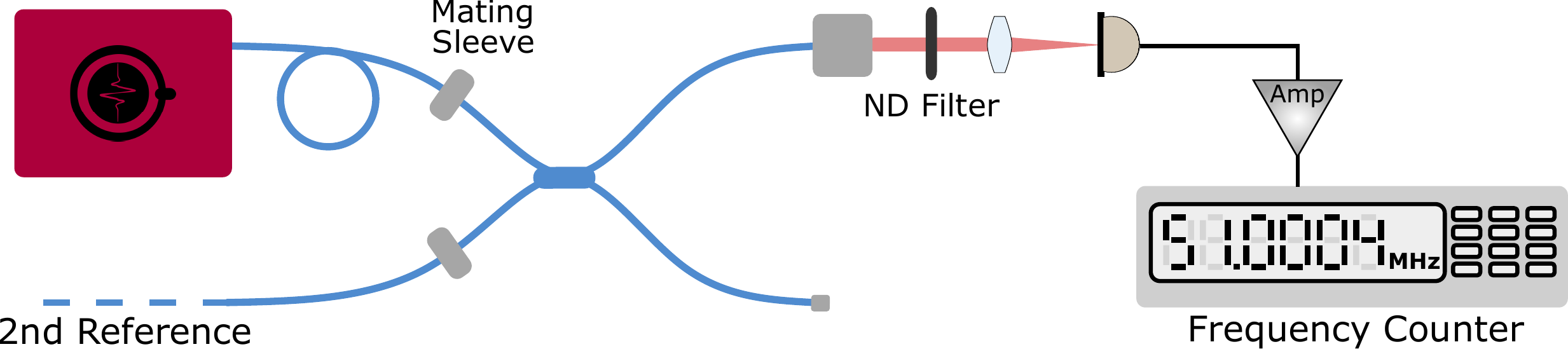}
\caption{Illustration of the beat-note detection setup.}
\label{fig:beat-note}
\end{figure}

\begin{figure}[ht]
\centering
\includegraphics[width=21pc]{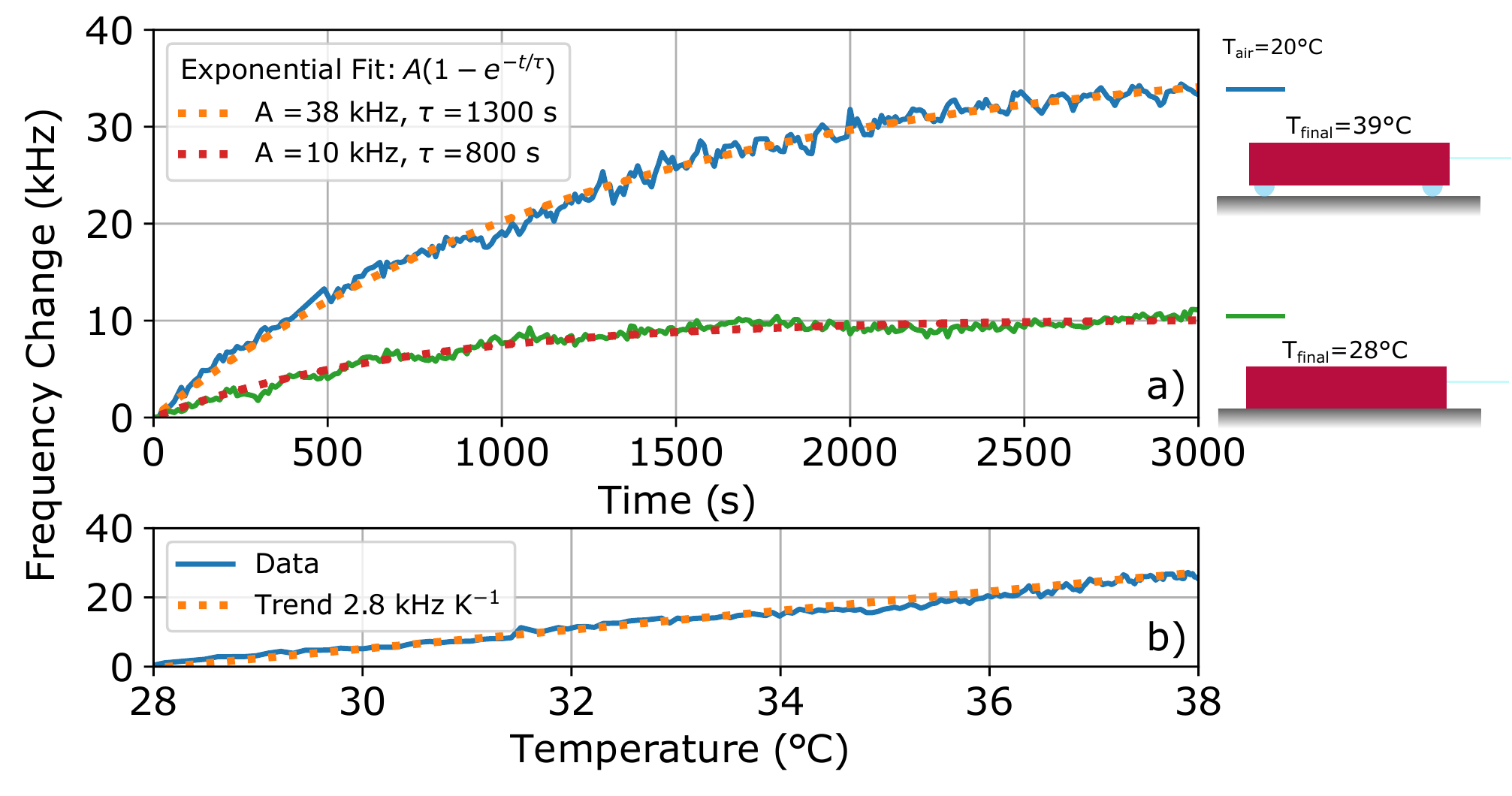}
\caption{Warm-up process of the system. (a) A frequency deviation of 4\,kHz, corresponding to a relative deviation of 10$^{-11}$, from the final frequency value is reached after about 3000\,s, when the system is mounted on rubber stands (green). When in direct contact with the optical table, this level is reached after about 700\,s. (b) Temperature dependence of the beat-note frequency. The dependence of 2.8\,kHz K$^{-1}$ corresponds to a relative frequency deviation of 7.3$\times$10$^{-12}$ K$^{-1}$.}
\label{fig:warmup}
\end{figure}

\begin{figure}[ht]
\centering%
\includegraphics[width=18pc]{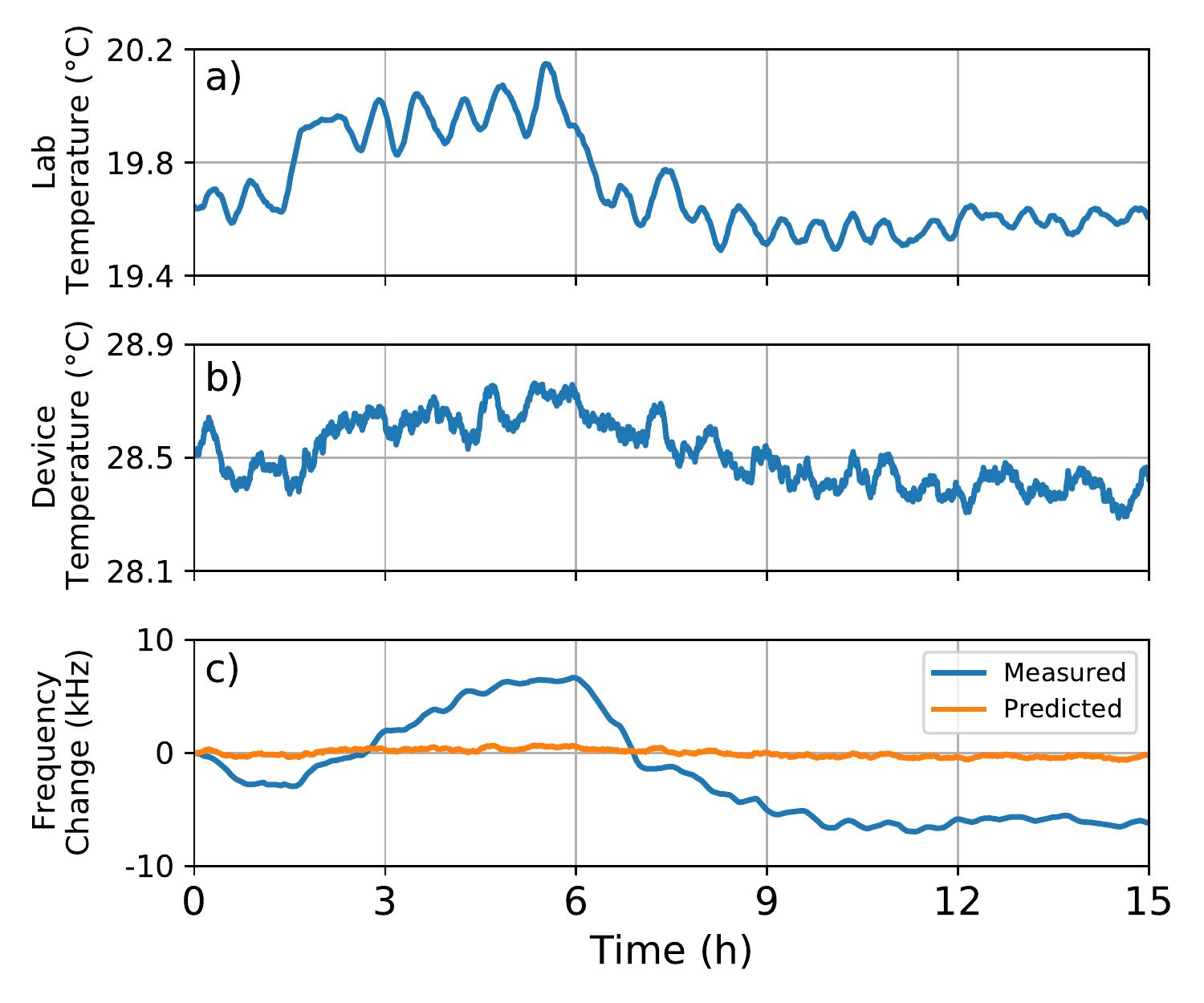}
\caption{Measurement of the impact of temperature changes in the laboratory. After 2\,h the temperature rises from 19.7\,$^{\circ}$C to 20\,$^{\circ}$C and sinks down to 19.5\,$^{\circ}$C after 6\,h. A clear correlation between the laboratory temperature (a), the device temperature (b) and the beat-note frequency (c) can be identified. The predicted beat-note frequency (orange) is calculated using the temperature coefficient from Figure \ref{fig:warmup} based on the device temperature. The frequency variations are too small by a factor of 10 in comparison to the measured ones (blue). This observation is suspected to be attributable to the second reference laser or components in our beat-note setup.}
\label{fig:tempsensgain}\vspace*{-6pt}
\end{figure}

The system is passively cooled by its casing. With a constant ambient temperature, the device temperature during the warm-up process $T(t)$ can be modelled by Newton's law of cooling \cite{Newton}:
\begin{equation}\label{eq:warm_up}
    \Delta T(t)=A(1-e^{-\frac{t}{\tau}}),
\end{equation}
where $\Delta T(t)$ is defined as $\Delta T(t)=T(t)-T(0)$, $A=T_{\mathrm{final}}-T(0)$ is the maximum temperature change and $\tau$ is a time constant defined by the heat capacitance and thermal conductance. To characterize the warm-up process in regard to its effect on the optical frequency, the latter is measured as soon as the laser is locked, which happens approximately 2 minutes after the cold start. In case of a warmed-up system, ramping up the laser current, stabilizing the laser diode temperature and initializing the lock does only take 20\,s. Currently, this time is limited by the manually defined minimum settling time for the temperature (10\,s) and can be optimized if needed. A beat-note measurement with a second rubidium-based reference \cite{Gain} utilizing MTS serves for measuring the frequency change. Figure \ref{fig:beat-note} shows the measurement setup.
In Figure \ref{fig:warmup} a) we analyze the absolute frequency change over time with equation \ref{eq:warm_up}. Two scenarios are shown: the blue line shows a measurement where the system is isolated from the optical table by the rubber stands. The green line shows the measurement with the reference being in contact with the optical table. Both measurements happened at an ambient temperature of 20\,$^{\circ}$C. In the first case, the device temperature approaches 39\,$^{\circ}$C and the optical frequency settles after about 50 minutes. Under these conditions the internal temperature of the RP reaches 48\,$^{\circ}$C, which is well below  85\,$^{\circ}$C, the recommended maximum operating temperature of the \textit{Xilinx} field-programmable gate-array \cite{Xilinx}. With the stands removed, the temperature settles at 28\,$^{\circ}$C after about 12 minutes. Since the changes in optical frequency in both cases are in the range of tens of kHz, the AOFR can be used for rubidium cold atom applications 2 minutes after cold start and immediately after locking from stand-by mode.\\ Figure \ref{fig:warmup} b) shows a linear relationship between the device temperature and the optical frequency of 2.8\,kHz K$^{-1}$. At an optical frequency of 384\,THz, this corresponds to a relative frequency deviation of 7.3$\times$10$^{-12}$\,K$^{-1}$. Thus, we can approximate that frequency instabilities below  10$^{-12}$ require a temperature instability lower than 130\,mK.
\\
Figure \ref{fig:tempsensgain} shows the results of a long-term frequency stability measurement, where the laboratory temperature changed by more than 0.5\,K over the course of 15\,h. The frequency response in c) is stronger by a factor of 10 than expected from the measured device temperature. This might indicate that either the second reference or components in our beat-note detection setup, which are placed in the same laboratory, have lead to a higher temperature sensitivity and the measured long-term stability needs to be considered an upper limit.

\section{Optimization of the Allan Deviation}\label{sec:opti}
\begin{figure}[ht]
\centering%
\includegraphics[width=30pc]{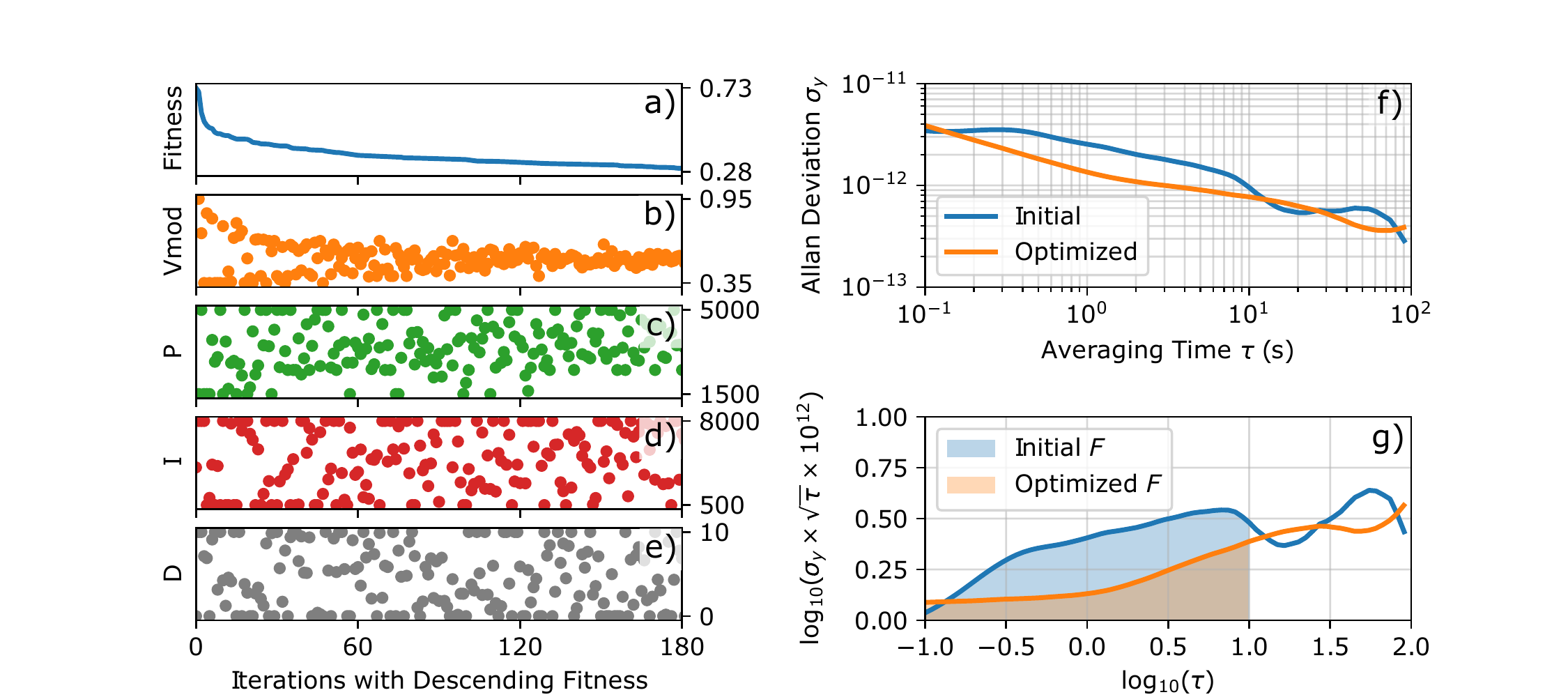}
\caption{Results of an exemplary Bayesian optimization process following initial manual optimization. (a) The steps are sorted based on descending fitness to visualize the convergence of the parameters. The strongest dependence is found on the modulation amplitude (b), while the PID parameters (c-e) show weak to no convergence. (f) Comparison of Allan deviation before and after the optimization process. The stability at 1\,s averaging time was improved by a factor of 2. (g) The fitness is illustrated by the shaded area below the two graphs. Deviations from a horizontal line indicate deviations from the $\tau^{-{1/2}}$ trend.}
\label{fig:optimization}\vspace*{-6pt}
\end{figure}

Initial optimization of the signal-to-noise ratio by increasing the modulation amplitude lead to the expected improvement of the short-term stability below 100\,ms averaging time. However, at longer averaging times, a deterioration from the $\tau^{-{1/2}}$ trend correlating with the modulation amplitude was observed. The Allan deviation for averaging times between 100\,ms and 10\,s is thus optimized with a Bayesian optimizer \cite{JCMWave} tuning the PID parameters of the lock and the modulation amplitude. The fitness function is chosen as:
\begin{equation}
    F=\int_{-1}^{1}\log_{10}\left(\sigma_y(\tau)\times\sqrt{\tau}\times10^{12}\right)\mathrm{d}\log_{10}(\tau).
\end{equation}
In the case of $\sigma_y(\tau)=a\frac{10^{-12}}{\sqrt{\tau}}$, $a$ is then given by $10^{F/2}$.\\
The beat-note frequency is recorded every 100\,ms for 600\,s in total, to reduce the statistical error up to 10\,s averaging time. Figure \ref{fig:optimization} a) shows the fitness and corresponding parameters sorted for descending fitness. The modulation amplitude converges to a value of 0.52$\pm$0.04\,V and the proportional parameter of the PID tends to 3700$\pm$800 towards the lowest fitness values. The two other parameters have no observable effect on the overall fitness. The initial parameters were set to Vmod=0.95\,V and P=5000. The optimized Allan deviation is reduced at times longer than 130\,ms on the cost of an increase at lower averaging times. This can be explained by a reduced signal-to-noise ratio due to a lower signal amplitude with lower modulation amplitude. The fitness is reduced from 0.73 to 0.28. The reason for the deterioration with increasing modulation amplitude is subject of ongoing investigations and might be the result of the side-band amplitude affecting the measurement with the frequency-counter.\\

\section{Performance}\label{sec:perform}
\begin{figure}[ht]
\centering
\includegraphics[width=21pc]{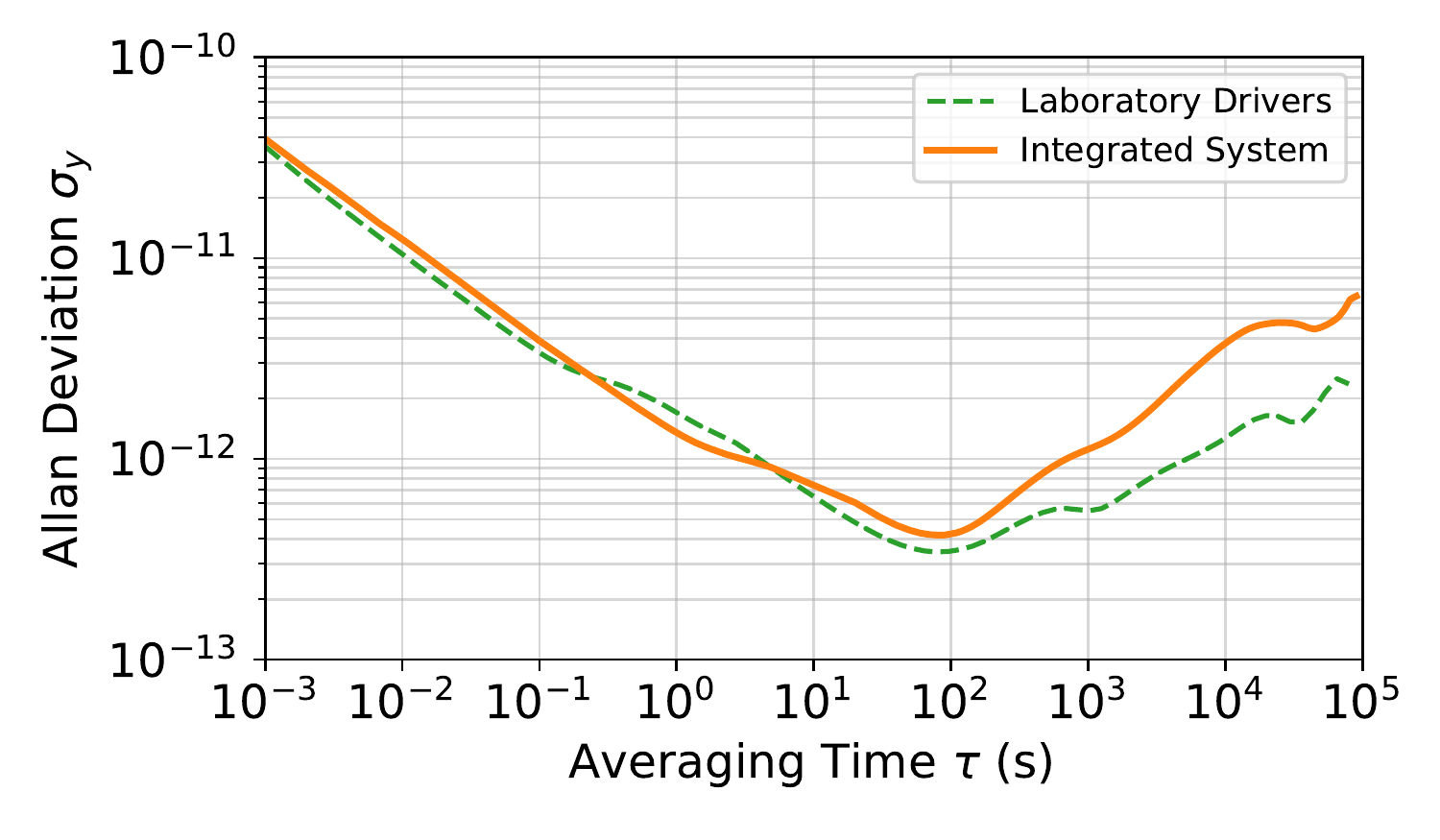}
\caption{Comparison of the Allan deviation of the optical system driven by external laboratory drivers \cite{Strangfeld:21} and the fully integrated system. An increased deviation at long time scales might be the result of stronger temperature fluctuations in the laboratory.}
\label{fig:finaladev}
\end{figure}

Figure \ref{fig:finaladev} shows the frequency stability for averaging times from 10$^{-3}$\,s to 10$^{5}$\,s. 
The displayed Allan deviation is composed of three measurement series with different measurement times and sampling rates: 10\,s with 1\,kHz, 10$^{3}$\,s with 10\,Hz, and finally 2$\times$10$^{5}$\,s with 0.1\,Hz. For comparison, the Allan deviation from \cite{Strangfeld:21}, where the optical module was driven with external laboratory drivers, is shown. As a result of the optimization process described in Section \ref{sec:opti} the Allan deviation is reduced to 1.4$\times$10$^{-12}$ at 1\,s averaging time. Despite the integration of all functional subsystems into a hand-held device, we managed to keep the increase of the frequency instability below 15\,\% up to 100\,s averaging time. The increase of the long-term instability to a value of 7$\times$10$^{-12}$ at an averaging time of 10$^{5}$\,s is suspected to be due to a less stable laboratory temperature during the new measurement.

\section{Discussion}\label{sec:discussion}
We have presented a fully-autonomous integrated AOFR enabling user-friendly access to frequency stabilized light at 780\,nm with a relative frequency instability of 1.4$\times$10$^{-12}$ at 1\,s and below 10$^{-11}$ at 10$^5$\,s averaging time. A combination of adapted COTS components and custom electronics allowed for a compact, yet cost-efficient design. We have shown that under standard laboratory conditions the system is locked with initial relative frequency deviations of 2.6$\times$10$^{-11}$ after 2 minutes following a cold start. After a warm-up period of 12 minutes a minimum Allan deviation of 4$\times$10$^{-13}$ at 100 seconds averaging time is achieved. We could also demonstrate an exemplary application of a Bayesian optimizer to short-term frequency instabilities, where we achieved a reduction by up to a factor of 2.\\Both, power consumption and warm-up time can be expected to be further reduced by a custom control unit instead of the RP. The temperature sensitivity of 7.3$\times$10$^{-12}$\,K$^{-1}$ is likely to limit the long-term stability in most applications if not compensated for by an additional temperature stabilization of the casing. Under temporarily stable laboratory temperatures we observed a flicker noise floor at about 3$\times$10$^{-13}$, which would then limit the long-term stability under optimized thermal design. The device in its current state may be used for cold-atom experiments in laboratories or as a building block for future cold-atom quantum sensors and wavelength measurements or calibration.\\ Our platform can be adapted for referencing to other alkali-vapors like potassium \cite{Kalexus} or cesium \cite{CesiumMEMS}, but also for more advanced schemes. To achieve long-term stabilities relevant to optical clock applications for navigation, compact AOFRs based on the two-photon transition at 778\,nm are promising candidates \cite{2photon,2PhotonHighPerf}. Here, using MEMS vapor cells instead of glass-blown ones could improve the scalability of the device. Further size reduction is achieved by photonic integration of such vapor cells \cite{ChipScaleD2}. Intrinsically higher short-term stability compared to the presented D2-line based system could also be realized utilizing a GaN laser diode and the 5S-6P transition at 420nm in rubidium \cite{ref420nm}.

\begin{backmatter}
\bmsection{Funding}
The authors acknowledge support by the German Space Agency DLR with funds provided by the Federal Ministry for Economic Affairs and Energy (BMWi) under grant numbers 50RK1971 (ROSC) and 50WM2066 (OPTIMAL-QT), and by the Berlin University Alliance (BUA).

\bmsection{Acknowledgments}
The authors greatly thank the workshop at Humboldt-Universität zu Berlin for technical support and B. Leykauf for providing the second reference for the beat-note measurements.

\bmsection{Disclosures}
The authors declare no conflicts of interest.

\bmsection{Data availability}
Data underlying the results presented in this paper are not publicly available at this time but may be obtained from the authors upon reasonable request.

\end{backmatter}
%%%%%%%%%% If using BibTeX:
\bibliography{fruby}

%%%%%%%%%% If preparing manually:
% \newcommand{\enquote}[1]{``#1''}

% \bibitem{Zhang:14}
% Y.~Zhang, S.~Qiao, L.~Sun, Q.~W. Shi, W.~Huang, L.~Li, and Z.~Yang,
%   \enquote{Photoinduced active terahertz metamaterials with nanostructured
%   vanadium dioxide film deposited by sol-gel method,}
%   {\protect\JournalTitle{Optics Express}} \textbf{22}, 11070--11078 (2014).

% {Optical Society}, \enquote{{OSA Publishing},}
%   \url{http://www.osapublishing.org}.

% \bibitem{FORSTER2007}
% P.~Forster, V.~Ramaswamy, P.~Artaxo, T.~Bernsten, R.~Betts, D.~Fahey,
%   J.~Haywood, J.~Lean, D.~Lowe, G.~Myhre, J.~Nganga, R.~Prinn, G.~Raga,
%   M.~Schulz, and R.~V. Dorland, \enquote{Changes in atmospheric consituents and
%   in radiative forcing,} in \enquote{Climate Change 2007: The Physical Science
%   Basis. Contribution of Working Group 1 to the Fourth assesment report of
%   Intergovernmental Panel on Climate Change,}  S.~Solomon, D.~Qin, M.~Manning,
%   Z.~Chen, M.~Marquis, K.~B. Averyt, M.~Tignor, and H.~L. Miler, eds.
%   (Cambridge University Press, 2007).

% \end{thebibliography}

\end{document}